\documentclass[aps,prb,showpacs,twocolumn]{revtex4-1}
\usepackage{bm,color,amsmath,amssymb,mathrsfs,latexsym,graphicx,psfrag}
\newcommand{\im}{\mathrm{Im}}

\newcommand{\be}{\begin{equation}}
\newcommand{\ee}{\end{equation}}

\def\be{\begin{equation}}
\def\ee{\end{equation}}
\def\bea{\begin{eqnarray}}
\def\eea{\end{eqnarray}}

\def\C60{A$_x$C$_{60}$}

\def\HgCu3{HgCa$_2$Cu$_3$O$_{8+y}$}
\def\HgCu4{HgBa$_2$Ca$_3$Cu$_4$O$_{10+y}$}
\def\TlCu{Tl$_2$Ba$_2$CuO$_{6+\delta}$}
\def\TlCu3{Tl$_2$Ba$_2$Ca$_2$Cu$_3$O$_{10+y}$}
\def\TlCu4{Tl$_2$Ba$_2$Ca$_3$Cu$_4$O$_{12+y}$}

\def\BiCu3{Bi$_2$Sr$_2$Ca$_{2}$Cu$_3$O$_y$}

\def\8LSCO{La$_{1.88}$Sr$_{.12}$CuO$_4$}
\def\110LNSCO{La$_{1.5}$Nd$_{0.4}$Sr$_{0.1}$CuO$_{4}$}
\def\stage4LCO{La$_{2}$CuO$_{4+\delta}$}
\def\Y248{YBa$_2$Cu$_4$O$_8$}

\def\NbSe2{NbSe$_2$}
\def\TaSe2{TaSe$_2$}
\def\TiSe2{TiSe$_2$}

\newcommand{\bk}{{\mathbf k}}

\newcommand{\br}{{\mathbf r}}
\newcommand{\bq}{{\mathbf q}}

\begin{document}

\title{Theory of quasiparticle interference in mirror symmetric 2D systems and its application to surface states of topological crystalline insulators}
\author{Chen Fang$^{1,2}$, Matthew J. Gilbert$^{3,4}$, Su-Yang Xu $^{2}$, B. Andrei Bernevig$^2$, M. Zahid Hasan$^{2}$}
\affiliation{$^1$Department of Physics, University of Illinois, Urbana IL 61801-3080}
\affiliation{$^2$Department of Physics, Princeton University, Princeton NJ 08544}
\affiliation{$^3$Department of Electrical and Computer Engineering, University of Illinois, Urbana IL 61801}
\affiliation{$^4$Micro and Nanotechnology Laboratory, University of Illinois, Urbana IL 61801}

\date{\today}

\begin{abstract}

We study symmetry protected features in the quasiparticle interference (QPI) pattern of 2D systems with mirror symmetries and time-reversal symmetry, around a single static point impurity. We show that, in the Fourier transformed local density of states (FT-LDOS), $\rho(\bq,\omega)$, while the position of high intensity peaks generically depends on the geometric features of the iso-energy contour at energy $\omega$, the \emph{absence} of certain peaks is guaranteed by the opposite mirror eigenvalues of the two Bloch states that are (i) on the mirror symmetric lines in the Brillouin zone (BZ) and (ii) separated by scattering vector $\bq$. We apply the general result to the QPI on the $\langle{001}\rangle$-surface of topological crystalline insulator Pb$_{1-x}$Sn$_x$Te and predict all vanishing peaks in $\rho(\bq,\omega)$. The model-independent analysis is supported by numerical calculations using an effective four-band model derived from symmetry analysis.

\end{abstract}
\maketitle

\section{Introduction} 
\label{intro}
Quantum interference is one of the simplest yet most fundamental quantum mechanical phenomena. In condensed matter physics, the interference pattern of quasiparticles around a single impurity observed through scanning tunneling spectroscopy (STS) has been widely used for electronic structure characterization of unconventional states\cite{McElroy2003,Aynajian2012}. For instance, QPI has been applied to demonstrate the defining property of time-reversal symmetric (TRS) topological insulators\cite{roushan2009,Xue:2009,seo2010,Okada2011,Alpichshev2010,Alpichshev2011}: the \emph{absence} of back-scattering in the surface states. In fact, a generic relationship between the absence of certain scattering channels and the underlying symmetry group exists in any system. In this Letter, we elucidate such relations in 2D systems with mirror symmetries by analyzing properties of the Fourier transformed local density of states around a single static impurity of an arbitrary kind (potential, magnetic, dipolar, etc.). We show that while generically the positions of high intensity peaks of FT-LDOS at a given energy are determined by the shape of the iso-energy contour\cite{Bena2009,Liu2012}, peaks at $\rho(\bq,\omega)$ vanishes if the two Bloch states separated by $\bq$ on the contour have opposite mirror eigenvalues. Moreover, the inverse statement is also true: by identifying the absence of these peaks in a multi-band system, we know that the two states mirror eigenvalues of each band on mirror symmetric lines in BZ, thus partly revealing the orbital nature of these bands. (`Orbital' includes orbital and spin degrees of freedom.)

The surface states of the recently discovered topological crystalline insulators (TCI) make an example of a multi-orbital 2D metal. Among the intensive theoretical efforts on topological phases beyond those protected by time-reversal symmetry\cite{schnyder2008,Fu:2011,hughes2010inv,Turner:2012,Fu:2012,Fang:2011,Fang2012}, Hsieh \emph{et al} predicted\cite{Fu:2012}, via first principles calculation, that SnTe is a nontrivial insulator whose topology is protected by mirror symmetries, dubbed mirror symmetric topological crystalline insulator (MSTCI), characterized by robust surface states on the $\langle{001}\rangle$-plane which contains in the surface Brillouin zone (BZ) four Dirac points (two along $\bar{X}$-$\bar{\Gamma}$-$\bar{X}$ and two along $\bar{Y}$-$\bar{\Gamma}$-$\bar{Y}$). This prediction has been experimentally confirmed\cite{Xu2012,Dziawa2012,Tanaka2012} through the use of angle resolved photoemission spectroscopy (ARPES) in Pb$_{1-x}$Sn$_x$Te. Furthermore\cite{Xu2012,Tanaka2012}, as the binding energy increases away from the Dirac point binding energy, the surface states undergo a Lifshitz transition at a critical energy $E_c$ where the iso-energy contour changes from two separate ellipsoids to two concentric ellipsoids. Theoretical proposals of other types of TCI materials have also appeared\cite{Kargarian2013,Liu2013,Liu2013a}. The defining property of a MSTCI is that the surface Dirac point is protected by \emph{mirror symmetry}, i.e., the two degenerate states at each Dirac point have different mirror eigenvalues. In this paper, we show that it can be promptly identified by examining the vanishing intensities in the FT-LDOS.

To numerically verify the theory, we use symmetry analysis to derive a minimal four-band model that semi-quantitatively recovers the observed dispersion and the Lifshitz transition. We calculate the FT-LDOS around a point charge impurity and see complete intensity suppression at wavevectors connecting two states with opposite mirror eigenvalues, even when the scattering is \emph{not} forbidden by TRS.

The paper is organized as follows. In Sec.\ref{general}, we develop a general analysis for the QPI in 2D systems with mirror symmetries and TRS, focusing on the general properties of the pattern and the symmetry forbidden scattering channels in FT-LDOS. In Sec.\ref{application}, we apply the general theory to the surface states of TCI SnTe, deducing all forbidden channels of FT-LDOS based on experimental facts only; then we develop an effective theory that captures the main features of the surface states and perform a numerical study which confirms our results. In Sec.\ref{discussion}, we discuss how the general theory can be applied to systems without spin orbital coupling, where TRS forbidden channels no longer exist and the limitations of our study.

\section{QPI for 2D systems with mirror symmetries and TRS} 
\label{general}
We begin by considering a multi-band homogeneous 2D system with the following symmetries: mirror reflections about $xz$- and $yz$-planes and TRS. In matrix representation, the single particle Hamiltonian $H(\bk)$ and symmetry operators satisfy: $M_{xz}H(k_x,k_y)M^{-1}_{xz}=H(k_x,-k_y)$, $M_{yz}H(k_x,k_y)M^{-1}_{yz}=H(-k_x,k_y)$, $U_TH(k_x,k_y)U_T^{-1}=H^T(-k_x,-k_y)$, where $H(\bk)$ is written in an arbitrarily chosen orbital basis and $M_{xz}$, $M_{yz}$ and $U_T$ ($T=KU_T$, with $K$ being complex conjugation) are the unitary matrices representing the symmetry operations in the same basis. In the presence of impurities, the local density of states is defined as $\rho(\br,\omega)=\sum_M|\langle{\br}|\psi_M\rangle|^2\delta(\omega-E_M)$, where $|\psi_M\rangle$ is the $M$-th single particle eigenstate and $E_M$ the corresponding energy. Under Fourier transform, it becomes the FT-LDOS
\bea\label{eq:rho}
\nonumber&&\rho(\bq,\omega)=\\
&&-\frac{1}{N\pi}\sum_{\bk}Tr[G(\bk,\bk+\bq,\omega)-G^\ast(\bk+\bq,\bk,\omega)],
\eea
where $N$ is the total number of unit cells in the system, and $G(\bk,\bk+\bq,\omega)$ the retarded Green's function. For a single point impurity given by a matrix $V$ in the chosen basis, the exact T-matrix approach solves $G(\bk,\bk+\bq,\omega)$ in terms of the homogeneous Green's function $G_0(\bk,\omega)$:
\bea\label{eq:fullG}
\nonumber&&G(\bk,\bk+\bq,\omega)=\\
&&G_0(\bk,\omega)\delta_{\bq0}+G_0(\bk,\omega)T(\omega)G_0(\bk+\bq,\omega),
\eea
where $G_0(\bk,\omega)=(\omega-H(\bk)+i\eta)^{-1}$. The matrix $T(\omega)$ contains the effect of impurity and is defined as
\bea
T(\omega)=V(I-g(\omega)V)^{-1},
\eea
where $g(\omega)=\sum_{\bk}G_0(\bk,\omega)/N$. The pattern of $\rho(\bq,\omega)$ naturally depends on the type of impurity, namely, the form of matrix $V$. $V$ is denoted by three indices $(\eta_x,\eta_y,\eta_T)$, where $\eta_{x,y,T}=\pm1$ means that the impurity changes/does not change sign under $M_{xz,yz}$ and TRS, respectively. For example, the combination $\eta=(1,-1,1)$ is an impurity that is invariant under mirror reflection about $xz$-plane and TRS yet changes sign under a mirror reflection about $yz$-plane. The physical realization of this type is an electric dipole oriented along $x$-axis. For any of the eight types, we derive the exact and approximate symmetries of $\rho(\bq,\omega)$ and summarize them in Table \ref{tab:impurity} (for derivation, see Appendix \ref{derivation}). Here, `approximate' indicates that the relation is valid only when $N_0(\omega)||V||\ll1$ where $N_0$ is the homogeneous density of states at $\omega$.

\begin{table*}
\includegraphics[width=12cm]{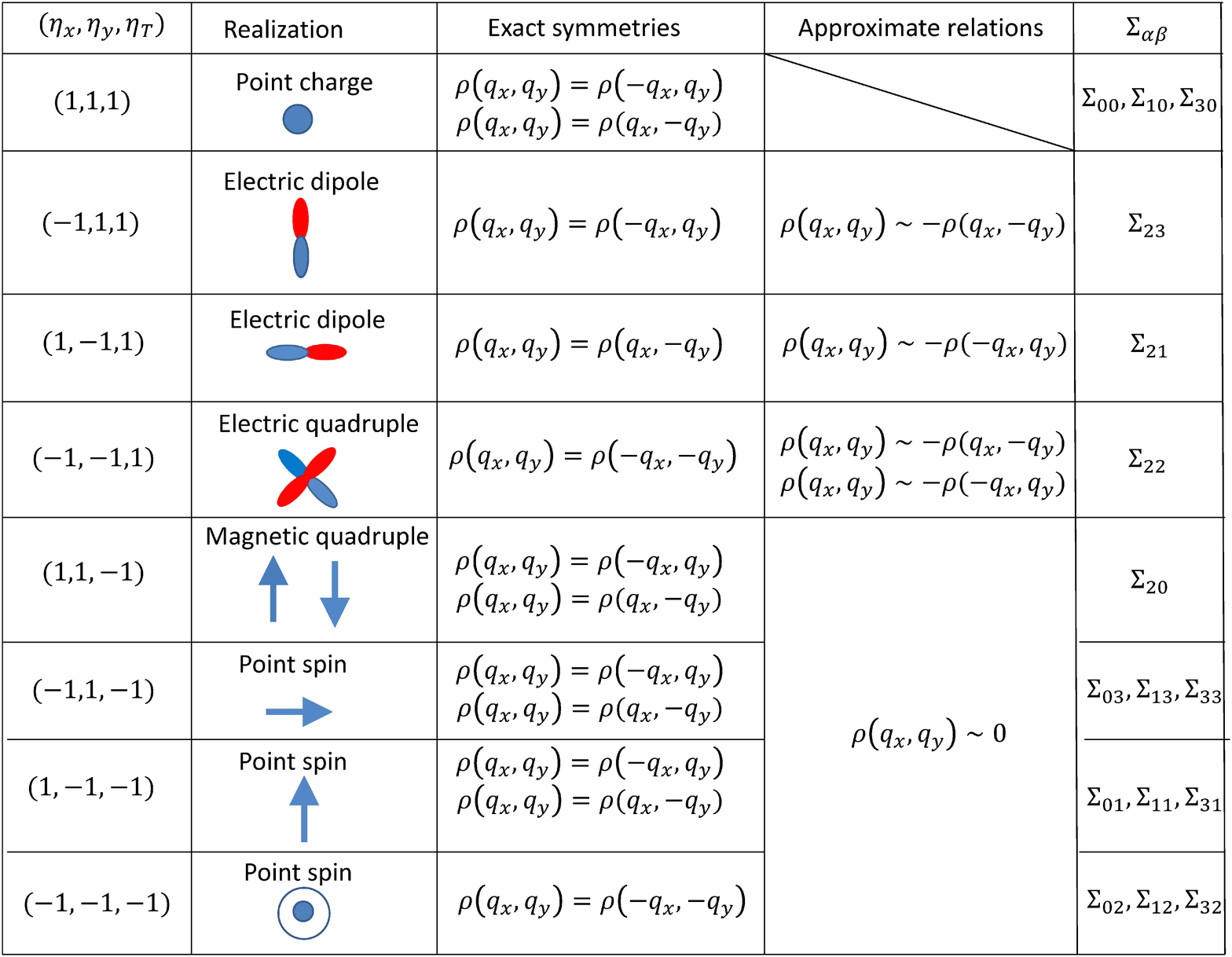}
\caption{Different types of point impurities classified by whether they change signs under mirror symmetries and TRS. From the left: the $\eta$-index, the possible physical realization, the exact symmetries of $\rho(\bq,\omega)$, the approximate symmetries of $\rho(\bq,\omega)$ to the first order of $V$ and the corresponding $\Sigma_{\alpha\beta}$ in the four-band effective model (see text).}
\label{tab:impurity}
\end{table*}

Besides the global symmetries in $\rho(\bq,\omega)$ derived above, the most important feature of the FT-LDOS is the presence of \emph{singularities} at certain scattering vectors, i.e., $\bq$-vectors. We start from Eq.(\ref{eq:fullG}) by rewriting the retarded Green's function in terms of projectors:
\bea\label{eq:G_projector}
\nonumber&&Tr[G(\bk,\bk+\bq,\omega)]=\\&&\sum_{mn}\frac{Tr[P_n(\bk+\bq)P_m(\bk)T(\omega)]}{(\omega-E_m(\bk)+i\eta)(\omega-E_n(\bk+\bq)+i\eta)},
\eea
where $n$ is the band index and $P_n=|u_n(\bk)\rangle\langle{u}_n(\bk)|$ is the projector onto the $n$-th band, and $|u_n(\bk)\rangle$ is the periodic part of the Bloch state at $\bk$ with band index $n$. (In Eq.(\ref{eq:G_projector}), we ignore the  first term in Eq.(\ref{eq:fullG}) that gives the homogeneous DOS, as we are interested in the inhomogeneous part of LDOS.) The most significant contribution to $\rho(\bq,\omega)$ comes from regions near the two poles of Eq.(\ref{eq:fullG}) at $\omega=E_m(\bk)$ and $\omega=E_n(\bk+\bq)$ respectively. Physically, $\rho(\bq,\omega)$ comes from the elastic scattering between two states of energy $\omega$ possessing a momentum difference $\bk_1-\bk_2=\bq$. From this picture, we know that the FT-LDOS is nonzero at $(\bq,\omega)$ if there are two states on the iso-energy contour of $\omega$ and separated by $\bq$. In fact, closer study\cite{Biswas2010,Liu2012} shows that $\rho(\bq,\omega)$ is divergent at $\bq$, if the normal directions, of the contour at $\bk_1$ and $\bk_2$, determined by the gradient of the dispersion, are parallel or anti-parallel. As the gradient of dispersion is the quasiparticle group velocity, this shows that the forward (with zero scattering angle) and backward scattering (with $\pi$ scattering angle) are the major channels. Any pair of $\bk_{1,2}$'s satisfying this condition are called a pair of stationary points denoted by $\bk_{s1,s2}$, and generically $\rho(\bq,\omega)$ is divergent at $\rho(\bk_{s1}-\bk_{s2},\omega)$. Each divergence in reality becomes a finite peak due to finite quasiparticle lifetime [finite $\eta$ in Eq.(\ref{eq:G_projector})]. The intensity of the peak is given by the pre-factor $Tr[P_n(\bk_{s2})P_m(\bk_{s1})T(\omega)]$ [see Eq.(\ref{eq:G_projector})]. When the two states at $\bk_{s1,s2}$ are \emph{orthogonal}, we have $Tr[P_n(\bk_{s2})P_m(\bk_{s1})T(\omega)]=0$ hence the peak at $\rho(\bk_{s1}-\bk_{s2},\omega)\sim0$ is strongly suppressed. This \emph{absence} of singularities in the FT-LDOS pattern is a symmetry protected property of the system, because in general $\langle{u}_m(\bk_{s1})|u_n(\bk_{s2})\rangle=0$ can only be guaranteed by the presence of an underlying symmetry. For example in TRS TI, the two states at $\bk$ and $-\bk$ on the iso-energy contour are orthogonal because they are related by TRS. This leads to the suppression of $\rho(2\bk,\omega)$ that is a hallmark of TRS TI\cite{ZhouX2009,Lee:2009}. Mark that the suppression discussed here is completely independent of the symmetries of impurity, but only depends on the symmetries of the homogeneous system.

In a system with mirror symmetries $M_{xz}$ and $M_{yz}$, $M_{xz}$ ($M_{yz}$) ensures the orthogonality between two states with \emph{different} eigenvalues of $M_{xz}$ ($M_{yz}$) thus leading to forbidden scattering channels. In general, $M_{xz}$ symmetry gives $E_n(k_x,k_y)=E_n(k_x,-k_y)$ and specially at $k_y=k_{inv}$ it gives $\partial_{k_y}E_n(k_x,k_{inv})=0$, where $k_{inv}=0\;\textrm{or}\;\pi$. Therefore, any two states on the iso-energy contour with $k_y=k_{inv}$ are a stationary pair. Additionally, as $[M_{xz},H(k_x,k_{inv})]=0$, these states are eigenstates of $M_{xz}$, having eigenvalue $\pm{i}$\cite{Fang2012}. The general conclusion is that if the states at $(k_1,k_{inv})$ and $(k_2,k_{inv})$ have different mirror eigenvalues, the peak in FT-LDOS at $(\pm(k_1-k_2),k_{inv})$ is suppressed. The inverse statement is also true: if FT-LDOS due to stationary pair scattering at $(q,0)$ suppressed, where $q=k_1-k_2$ is the scattering vector between two states along $k_y=k_{inv}$, then states at $(k_1,k_{inv})$ and $(k_2,k_{inv})$ have opposite eigenvalues of $M_{xz}$. Similar statements can be made for the states along $k_y$ axis on the iso-energy contour at $k_x=k_{inv}$, using $M_{yz}$ symmetry.

Experimentally, the suppressed peaks in FT-LDOS is observed by comparing it to the joint density of states (JDOS), $\rho_J(\bq,\omega)$, or, the number of states per unit energy at energy $\omega$ and momenta $\bk_{1,2}$ satisfying $\bk_1-\bk_2=\bq$. The JDOS can be experimentally obtained by convolving the ARPES measured spectral weight\cite{roushan2009}. Formally, $\rho_J(\bq,\omega)$ is defined as
\bea
&&\rho_J(\bq,\omega)=\\
&&\nonumber-\frac{2}{N\pi}\im[\sum_{m,n,\bk}\frac{1}{(\omega-E_m(\bk)+i\eta)(\omega-E_n(\bk+\bq)+i\eta)}].
\eea
When compared with Eq.(\ref{eq:G_projector}), the only difference is that in $\rho_J(\bq,\omega)$ the factor $Tr[P_m(\bk)T(\omega)P_n(\bk+\bq)]$ is missing. Hence $\rho_J(\bq,\omega)$ is completely determined by the geometric features of the iso-energy contour. A comparison between $\rho(\bq,\omega)$ and $\rho_J(\bq,\omega)$ thus gives information of the wave functions, because any stationary vector that is present in $\rho_J(\bq,\omega)$ but absent in $\rho(\bq,\omega)$ indicates that the two wave functions at the contributing stationary points are orthogonal.

\section{An application to identifying MSTCI} 
\label{application}
\subsection{Absent peaks in FT-LDOS}

\begin{figure}[!hbt]
\includegraphics[width=8cm]{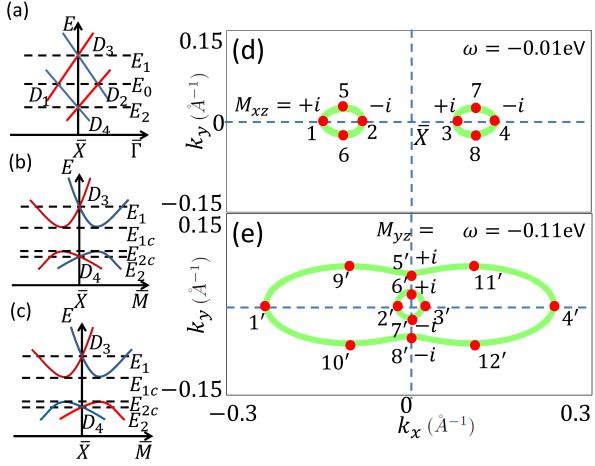}
\caption{Schematic band dispersion of the surface bands at vicinity of $\bar{X}$ along (a) $k_x$-axis and (b) $k_y$-axis. Red/blue (lighter/darker grey) means the band has eigenvalue $+i$/$-i$ of $M_{xz}$ in (a) and $M_{yz}$ in (b,c). Iso-energy contour calculated for the model described in the text at $\omega=-0.01$eV in (d) and $\omega=-0.11$eV in (e). Eigenvalues of $M_{xz}$ at points $1,2,3,4$ are marked beside each point, and eigenvalues of $M_{yz}$ at points $5',6',7',8'$ are similarly marked. The eigenvalues of $M_{xz}$ at points $1',2',3',4'$ are identical to those at $1,2,3,4$.}\label{fig:FS}
\end{figure}

We apply the general result to surface states of Pb$_{1-x}$Sn$_x$Te, to elucidate how its unique topological properties manifest themselves in QPI. The surface states exist around $\bar{X}$ and $\bar{Y}$\cite{Xu2012,Dziawa2012,Tanaka2012}, and let us first focus on the states around $\bar{X}$, then due to $C_4$ symmetry, a parallel analysis can be repeated for those around $\bar{Y}$. Along $\bar\Gamma\bar{X}\bar\Gamma$, there are two Dirac points (denoted by $D_{1,2}$), or, band crossings. Since that the two bands crossings have opposite eigenvalues of $M_{xz}$, and that TRS inverts $M_{xz}$ eigenvalue from $\pm{i}$ to $\mp{i}$, the only possible configuration of $M_{xz}$ eigenvalues is the one shown in Fig.\ref{fig:FS}(a), up to an overall sign. Besides the two Dirac points at Fermi energy $E_0$, there are two additional Dirac points higher and lower than $E_0$ at $\bar{X}$ (denoted by $D_{3,4}$). Along the perpendicular direction $\bar{M}\bar{X}\bar{M}$, however, there is no band crossing point. Along this line, bands can either have the same eigenvalues of $M_{yz}$ and hence repel each other (Fig.\ref{fig:FS}(b)), or they can have the same eigenvalue of $M_{yz}$ but opposite effective mass (Fig.\ref{fig:FS}(c)). $D_{3,4}$ are protected by TRS and therefore appear in both directions; the other two Dirac nodes, $D_{1,2}$ are protected only by $M_{xz}$ symmetry. To see the Lifshitz transition, we find that for $E_{2c}<\omega<E_{1c}$, the bands along $\bar{X}\bar{M}$ do not cross $\omega$, resulting in an iso-energy contour of two separate ellipsoids centered at $D_{1,2}$ respectively. From Fig.\ref{fig:FS}(a), along $\bar{X}\bar{\Gamma}$, the two points on the same side of $\bar{X}$ have opposite eigenvalues of $M_{xz}$. While for $\omega>E_{c1}$ or $\omega<E_{c2}$, the bands along $\bar{X}\bar{M}$ cross $\omega$ twice, resulting in an iso-energy contour of two concentric loops centered at $\bar{X}$. The two crossing points on same side of $\bar{X}$ have the same mirror eigenvalue according to Fig.\ref{fig:FS}(b/c). This analysis gives us the mirror eigenvalues on the iso-energy contours before and after the Lifshitz transition, marked in Fig.\ref{fig:FS}(d,e). We now use the general theory to obtain the following pairs of suppressed peaks in FT-LDOS: $\bq_{12}$, $\bq_{34}$ and $\bq_{23}$ before the Lifshitz transition ($\bq_{ab}$ defined as $\bq_a-\bq_b$), and $\bq_{1'2'}$, $\bq_{3'4'}$, $\bq_{2'3'}$, $\bq_{5'7'}$, $\bq_{6'7'}$ and $\bq_{5'8'}$ after the Lifshitz transition. We note that $\bq_{14}$, $\bq_{23}$, $\bq_{1'4'}$, $\bq_{2'3'}$, $\bq_{5'8'}$, $\bq_{6'7'}$ are also forbidden by TRS as they are scattering vectors connecting Kramer's pairs in the presence of TRS.

\subsection{Effective model and numerics} 

Thus far, we have extracted the symmetries of the single impurity QPI pattern and located all vanishing singularities guaranteed by the presence of mirror symmetries, as one expects in Pb$_{1-x}$Sn$_x$Te. In order to be more concrete, we develop a minimal $\bk\cdot\mathbf{p}$-model which qualitatively captures the surface band dispersion both before and after the Lifshitz transition and calculate $\rho(\bq,\omega)$ around a point impurity (most common impurity in Pb$_{1-x}$Sn$_x$Te). Near the Dirac points $D_{1,2}$, the system can be approximated by two separate two-band $\bk\cdot\mathbf{p}$ models. Nevertheless, in order to describe the Lifshitz transition where the two cones cross each other, one needs to consider the hybridization between the two cones and, thus, the minimal model is at least four-band. At $\bar{X}$ there are two degenerate states at $D_3$ with $E_1>E_0$ and two other degenerate states at $D_4$ with $E_2<E_0$. Within each doublet, one is an eigenstate of $M_{yz}$ with an eigenvalue of $+i$ while the other $-i$. We define our basis as: $(1,0,0,0)$ and $(0,1,0,0)^T$ have energy $E_1$ and mirror eigenvalues of $M_{yz}=+i$ and $M_{yz}=-i$ respectively; $(0,0,1,0)^T$ and $(0,0,0,1)^T$ have energy $E_2$ and mirror eigenvalues of $M_{yz}=+i$ and $M_{yz}=-i$ respectively. There is gauge freedom in choosing the matrices for $M_{xz}$ and $U_T$ under this basis, and we choose $M_{xz}=i\Sigma_{01}$ and $U_T=i\Sigma_{02}$, where $\Sigma_{\alpha\beta}=\sigma_\alpha\otimes\sigma_\beta$. Applying symmetry constraints, we find that, to the linear order in $\bk$, $\Sigma_{30}$ and $\Sigma_{10}$ couple to the zeroth order, $\Sigma_{01,11,31}$ couple to $k_x$ and $\Sigma_{03,13,33}$ couple to $k_y$. The effective Hamiltonian can hence be written as
\bea\nonumber H(\bk)&=&m\Sigma_{30}+m'\Sigma_{10}+(v_{1x}\Sigma_{01}+v_{2x}\Sigma_{11}+v_{3x}\Sigma_{31})k_x\\
&+&(v_{1y}\Sigma_{03}+v_{2y}\Sigma_{13}+v_{3y}\Sigma_{33})k_y.\eea 
Qualitatively fitting the data for Pb$_{1-x}$Sn$_x$Te given in Ref.[\onlinecite{Xu2012}], we choose the parameters in our Hamiltonian to be: $\{m,m'\}=\{-0.11,-0.1\}$eV, $\{v_{1x},v_{2x},v_{3y}\}=\{-1.5,-0.5,-3\}$eV$\cdot\AA$ and $\{v_{3x},v_{1y},v_{2y}\}\sim0$. Here we define $\sim0$ as being very small but nonzero in order to avoid accidental symmetries. We represent $V$ for impurities with different $\eta$-indices in a similar fashion using this basis, given in the last column of Table \ref{tab:impurity}.

\begin{figure}[!hbt]
\includegraphics[width=7cm]{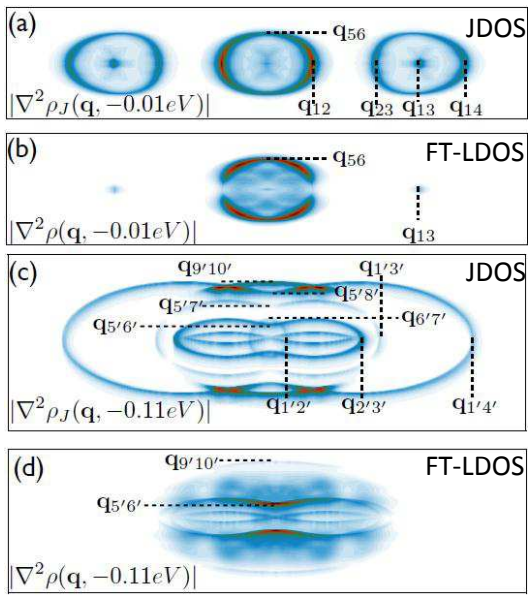}
\caption{The absolute value of the divergence of the joint density of states, $|\nabla^2\rho_J(\bq,\omega)|$, is plotted for $\omega=-0.01$eV (a) and $-0.11$eV (c). The absolute value of the divergence of FT-LDOS around a static impurity potential $V=0.1eV*\Sigma_{00}$, $|\nabla^2\rho(\bq,\omega)|$, is plotted for $\omega=-0.01$eV (b) and $-0.11$eV (d).}\label{fig:rho}
\end{figure}

Using the effective Hamiltonian for Pb$_{1-x}$Sn$_x$Te, we calculate $\rho(\bq,\omega)$ with $V=0.1eV*\Sigma_{00}$, i.e., a static point charge impurity. In Fig.\ref{fig:rho}(b,d), we plot $|\nabla^2\rho(\bq,\omega)|$ at $\omega=-0.01eV$ and $\omega=-0.11eV$ respectively. ($|\nabla^2\rho(\bq,\omega)|$ is plotted in lieu of $\rho(\bq,\omega)$ since it shows the singularities more clearly.) In comparison, we make the same plots for JDOS, $|\nabla^2\rho_J(\bq,\omega)|$, in Fig.\ref{fig:rho}(a, c). In Fig.\ref{fig:rho}(a, c), we mark several $\bq_s$'s that join stationary $\bk$-points marked in Fig.\ref{fig:FS}, and at all these vectors, there are peaks in $|\nabla^2\rho_J(\bq,\omega)|$. In contrast, in Fig.\ref{fig:rho}(b,d), we see that all peaks that are forbidden by either TRS or mirror symmetries vanish, confirming the previous analysis based on mirror eigenvalues of the $\bk$-points.

\subsection{QPI with rhombohedral distortion}

\begin{figure*}[!htb]
\includegraphics[width=16cm]{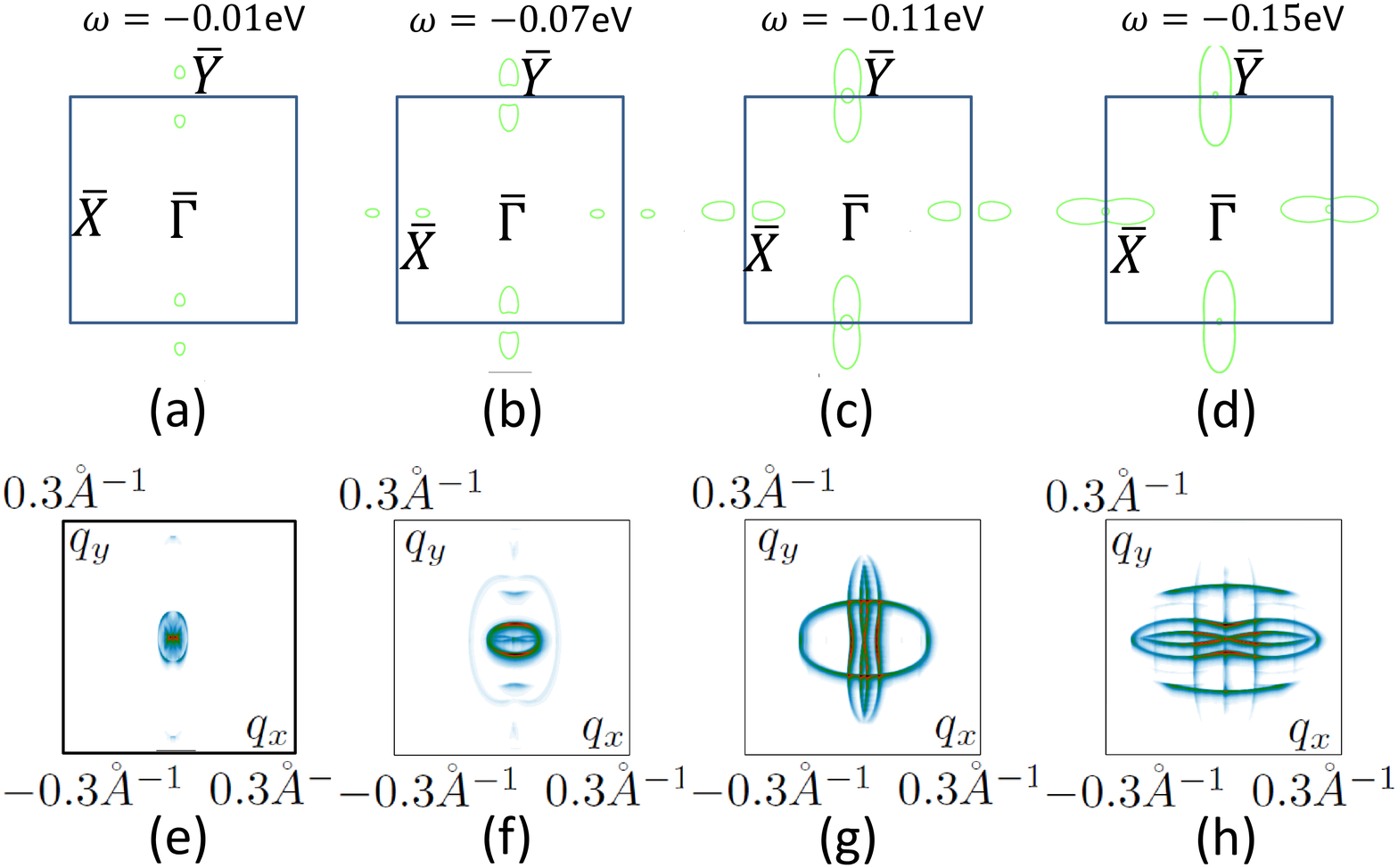}
\caption{(a,b,c,d): The iso-energy contours at $\omega=-0.01$eV, $-0.07$eV, $-0.11$eV and $-0.15$eV, respectively, in the phase with rhombohedral distortion, breaking $M_{xz}$ symmetry. (e,f,g,h): $\nabla^2\rho(\bq,\omega)$ calculated for he surface states in the rhombohedral phase around a single charge impurity at the same energies listed above.}\label{fig:gapped}
\end{figure*}

In samples with $x\sim1$ or pure SnTe, a spontaneous lattice distortion break one of the mirror symmetries (assumed to be $M_{yz}$) and $C_4$ symmetry. Assuming that the distortion breaks $M_{xz}$, we know that: first, the two Dirac nodes along $\bar{\Gamma}\bar{X}$ are no longer protected and open a gap of $2E_D$; second, the Lifshitz transition happens at two different energies, $E_{cx}$ and $E_{cy}$, for the surface states around $\bar{X}$ and around $\bar{Y}$. As energy decreases from $E_f$, there are four phases: (i) no FS around $\bar{X}$ and two ellipsoids around $\bar{X}$ when $|\omega|<E_D$, (ii) two ellipsoids of FS around $\bar{X}$ and two other around $\bar{Y}$ when $E_D<|\omega|<|E_{cy}|$, (iii) two ellipsoids around $\bar{X}$ and two concentric loops around $\bar{Y}$ if $|E_{cy}|<|\omega|<|E_{cx}|$ and (iv) two concentric loops around $\bar{X}$ and two others around $\bar{Y}$ if $|\omega|>|E_{cx}$. All these phases can be identified by QPI. In the pattern, all forbidden scattering vectors forbidden by $M_{xz}$ are present while those forbidden by $M_{yz}$ or TRS remain absent.

In the presence of the rhombohedral distortion, the fourfold rotation symmetry is broken and we have to consider the contribution from states near $\bar{X}$ and $\bar{Y}$ separately. Any term that breaks $M_{xz}$ but preserves $M_{yz}$ and TRS is of index $\eta=(-1,+1,+1)$. According to Table I, in the effective theory around $\bar{X}$, this term is $\Delta\Sigma_{23}$. Without distortion, the effective theory around $\bar{Y}$ is obtained by rotating the model in Eq.(6) by $\pi/2$, and in this theory, the distortion term corresponds to $\Delta\Sigma_{23}$.
In Fig.\ref{fig:gapped}, we choose $\Delta=0.1$eV and calculate the iso-energy contours at $\omega=-0.01$eV in (a), $-0.07$eV in (b), $-0.11$eV in (c) and $-0.15$eV in (d). From these figures, the change in the topology of the contours is clear: it changes from only two separate ellipsoids around $\bar{Y}$ in (a), to two separate ellipsoids around both $\bar{X}$ and $\bar{Y}$ in (b), to two separate ellipsoids around $\bar{X}$ and two concentric loops in (c), and finally to two concentric loops around both $\bar{X}$ and $\bar{Y}$ at highest $\omega$ in (d).
The resultant QPI patterns, $|\nabla^2\rho(\bq,\omega)|$, at these energies are plotted in Fig.\ref{fig:gapped}(e, f, g, h) respectively. These patterns are simply superpositions of patterns contributed by states around $\bar{X}$ and those around $\bar{Y}$; each pattern explicitly breaks fourfold rotational symmetry. Another important feature is that all $\bq$-vectors that are forbidden by $M_{xz}$ are now present. For example, in Fig.\ref{fig:gapped}(f), the inner circle is contributed by the surface states around $\bar{X}$. On this circle, the left and right ends, defined as $\bq_L$ and $\bq_R$ are due to the scattering between $\bk$'s at the two ends of either one of the two pockets near $\bar{X}$. When $M_{xz}$ is unbroken, these two $\bk$'s have opposite $M_{xz}$ eigenvalues and hence scattering between them is forbidden. But with $M_{xz}$ broken, $\bq_{L,R}$ are allowed by symmetry. Compare this with Fig.1(b), we can see that the distortion makes the intensity at $\bq_{L,R}$ finite.

\section{Discussion} 
\label{discussion}
The main results apply to any 2D or quasi-2D systems with mirror symmetries and, in a multi-orbital system can be used to identify the orbital nature of the bands on an iso-energy contour. For example, using our theory in conjunction with STS, one may use QPI to establish or disprove the long predicted Dirac metal state\cite{Ran2009,Hasan2010} in the SDW phase of 1111-family of iron-based superconductors. When applying the theory to the cases where spin-orbital coupling is negligible (e.g., in iron-based superconductors), or more accurately, where SU(2) symmetry is restored, we note the following differences: (i) mirror eigenvalues become real numbers $\pm1$ and (ii) there is no TRS forbidden channel due to spin degeneracy. Compared with ARPES with polarized light that also resolves orbitals for bands lower than the Fermi energy in the absence of magnetic field, QPI can be applied to field induced phases and to states above the Fermi energy.

In the application to Pb$_{1-x}$Sn$_x$Te, it is necessary to point out that we have assumed that $\rho(\bq,\omega)$ contains contributions only from scattering between states near $\bar{X}$. However, the full QPI pattern also consists of the scattering among states near $\bar{Y}$ and the inter-pocket scattering between states near $\bar{X}$ and those near $\bar{Y}$. Yet due to $C_4$-symmetry, the contribution to $\rho(\bq,\omega)$ made by states near $\bar{Y}$ can be obtained easily by making a fourfold rotation of $\rho(\bq,\omega)$ contributed by states around $\bar{X}$. The inter-scattering between states near $\bar{Y}$ and those near $\bar{X}$ only contributes to $\rho(\bq,\omega)$ near $\bq\sim\bar{X}-\bar{Y}=(\pi,\pi)$. As these features are sufficiently separated in $\bq$-space from those of in the small $\bq$-region, they are neglected in this work.

We have only studied the case of a single impurity, but the result extends to the case where there are random impurities of the same type and strength and when the impurity strength is weak. In Appendix \ref{autocorrelation},  we prove that under these conditions, the FT-LDOS around a single impurity, $\rho(\bq,\omega)$, can be related to the autocorrelation function of the overall LDOS resulted from many impurities, $R(\br,\omega)$.

\begin{acknowledgments}
CF is supported by ONR - N00014-11-1-0635 and N0014-11-1-0728. MJG acknowledges support from the AFOSR under grant FA9550-10-1-0459 and the ONR under grant N0014-11-1-0728. BAB was supported by NSF CAREER DMR- 095242, ONR - N00014-11-1-0635, Darpa - N66001-11- 1-4110, and David and Lucile Packard Foundation.
\end{acknowledgments}

\begin{appendix}
\onecolumngrid

\section{Derivation of the global symmetries of $\rho(\bq,\omega)$ for different types of single impurities}
\label{derivation}
In this section we explicitly derive the symmetries of $\rho(\bq,\omega)$, exact and approximate, listed in Table I of the main text for single impurities with eight possible $\eta$-indices.

From Eq.(1), $\rho(\bq,\omega)$ is determined by the retarded Green's function $G(\bk,\bk+\bq,\omega)$; yet from Eq.(2), $G(\bk,\bk+\bq,\omega)$ is given in terms of the free particle Green's function $G_0(\bk,\omega)$ and the T-matrix, $T(\omega)$. The former has exactly the same symmetries as the homogenous Hamiltonian:
\bea\label{eq:Gtransform}
M_{xz}G_0(k_x,k_y,\omega)M^{-1}_{xz}&=&G_0(k_x,-k_y,\omega),\\
\nonumber M_{yz}G_0(k_x,k_y,\omega)M^{-1}_{yz}&=&G_0(-k_x,k_y,\omega),\\
\nonumber U_TG_0(\bk,\omega)U_T^{-1}&=&G_0^T(-\bk,\omega).\eea
The symmetries of $T(\omega)$, however, depends on symmetries of both the system \emph{and} the impurity, i.e., the $\eta$-index. Using definitions of $\eta$-index and Eq.(3), one obtains:
\bea\label{eq:Ttransform1}
M_{xz}T(\omega)M_{xz}^{-1}=\eta_xV(I-\eta_xg(\omega)V)^{-1}&=&T(\omega)\;\textrm{if}\;\eta_x=1,\\
\nonumber                           								 &\sim&-T(\omega)\;\textrm{if}\;\eta_x=-1,\\
\nonumber M_{yz}T(\omega)M_{yz}^{-1}=\eta_yV(I-\eta_yg(\omega)V)^{-1}&=&T(\omega)\;\textrm{if}\;\eta_y=1,\\
\nonumber                           								 &\sim&-T(\omega)\;\textrm{if}\;\eta_y=-1,\\
\nonumber U_TT(\omega)U_T^{-1}=(\eta_TV(I-\eta_Tg(\omega)V)^{-1})^T&=&T^T(\omega)\;\textrm{if}\;\eta_T=1,\\
\nonumber                           								 &\sim&-T^T(\omega)\;\textrm{if}\;\eta_T=-1.
\eea
Here `$\sim$' means that the relation becomes exact only if $T(\omega)$ is replaced by its leading order, i.e., $T=V$.

Combining equations in Eqs.(\ref{eq:Ttransform1}), one easily obtains other two equations:
\bea\label{eq:Ttransform2}
(M_{xz}M_{yz})T(\omega)(M_{xz}M_{yz})^{-1}=\eta_x\eta_yV(I-\eta_x\eta_yg(\omega)V)^{-1}&=&T(\omega)\;\textrm{if}\;\eta_x\eta_y=1,\\
\nonumber                           								 &\sim&-T(\omega)\;\textrm{if}\;\eta_x\eta_y=-1,\\
(M_{xz}M_{yz}U_T)T(\omega)(M_{xz}M_{yz}U_T)^{-1}=[\eta_x\eta_y\eta_TV(I-\eta_x\eta_y\eta_Tg(\omega)V)^{-1}]^T&=&T^T(\omega)\;\textrm{if}\;\eta_x\eta_y\eta_T=1\\
\nonumber                           								 &\sim&-T^T(\omega)\;\textrm{if}\;\eta_x\eta_y\eta_T=-1.
\eea

Now we use Eqs.(\ref{eq:Gtransform}) and Eqs.(\ref{eq:Ttransform1}) and have
\bea
\rho(q_x,q_y,\omega)&=&\frac{1}{N\pi}\sum_\bk[Tr(M_{xz}G_0(k_x,k_y,\omega)T(\omega)G_0(k_x+q_x,k_y+q_y,\omega)M^{-1}_{xz})-(\bq\rightarrow-\bq)^\ast]\\
\nonumber&=&\frac{1}{iN\pi}\sum_\bk[Tr(G_0(k_x,k_y,\omega)M_{xz}T(\omega)M_{xz}^{-1}G_0(k_x+q_x,k_y-q_y,\omega))-(\bq\rightarrow-\bq)^\ast]\\
\nonumber&=&\rho(q_x,-q_y,\omega)\;\textrm{if}\;\eta_x=1\\
\nonumber&\sim&-\rho(q_x,-q_y,\omega)\;\textrm{if}\;\eta_x=-1,
\eea
\bea
\rho(q_x,q_y,\omega)&=&\frac{1}{iN\pi}\sum_\bk[Tr(M_{yz}G_0(k_x,k_y,\omega)T(\omega)G_0(k_x+q_x,k_y+q_y,\omega)M^{-1}_{yz})-(\bq\rightarrow-\bq)^\ast]\\
\nonumber&=&\frac{1}{iN\pi}\sum_\bk[Tr(G_0(k_x,k_y,\omega)M_{yz}T(\omega)M_{yz}^{-1}G_0(k_x-q_x,k_y+q_y,\omega))-(\bq\rightarrow-\bq)^\ast]\\
\nonumber&=&\rho(-q_x,q_y,\omega)\;\textrm{if}\;\eta_y=1\\
\nonumber&\sim&-\rho(-q_x,q_y,\omega)\;\textrm{if}\;\eta_y=-1,
\eea
\bea
\rho(q_x,q_y,\omega)&=&\frac{1}{iN\pi}\sum_\bk[Tr(U_TG_0(k_x,k_y,\omega)T(\omega)G_0(k_x+q_x,k_y+q_y,\omega)U_T^{-1})-(\bq\rightarrow-\bq)^\ast]\\
\nonumber&=&\frac{1}{iN\pi}\sum_\bk[Tr(G_0(-k_x-q_x,-k_y-q_y,\omega)(U_TT(\omega)U^{-1}_T)^TG_0(-k_x,-k_y,\omega))-(\bq\rightarrow-\bq)^\ast]\\
\nonumber&=&\frac{1}{iN\pi}\sum_\bk[Tr(G_0(k_x-q_x,k_y-q_y,\omega)(U_TT(\omega)U^{-1}_T)^TG_0(k_x,k_y,\omega))-(\bq\rightarrow-\bq)^\ast]\\
\nonumber&=&\rho(q_x,q_y,\omega)\;\textrm{if}\;\eta_T=1\\
\nonumber&\sim&-\rho(q_x,q_y,\omega)\;\textrm{if}\;\eta_T=-1.
\eea

We can also use Eqs.(\ref{eq:Gtransform}) and Eqs.(\ref{eq:Ttransform2}) to obtain, very similarly,
\bea
\rho(q_x,q_y,\omega)&=&\rho(-q_x,-q_y,\omega)\;\textrm{if}\;\eta_x\eta_y=1\\
\nonumber&\sim&-\rho(-q_x,-q_y,\omega)\;\textrm{if}\;\eta_x\eta_y=-1,\\
\rho(q_x,q_y,\omega)&=&\rho(-q_x,-q_y,\omega)\;\textrm{if}\;\eta_x\eta_y\eta_T=1\\
\nonumber&\sim&-\rho(-q_x,-q_y,\omega)\;\textrm{if}\;\eta_x\eta_y\eta_T=-1.
\eea
Up to now we have derived all symmetries, exact and approximate, of $\rho(\bq,\omega)$ listed in Table I.

\section{Autocorrelation of $R(\br,\omega)$ in the weak impurity limit}
\label{autocorrelation}
In the text, we mention that although we have only studied the case of a single impurity, the result extends to the case where there are random impurities of the same type and strength and when the impurity strength is weak. Here we prove that under these conditions, the FT-LDOS around a single impurity, $\rho(\bq,\omega)$, can be related to the autocorrelation function of the overall LDOS resulted from many impurities, $R(\br,\omega)$. By definition,
\bea
R(\br,\omega)=\sum_{n}|\psi_n(\br)|^2\delta(\omega-E_n)=-\frac{1}{\pi}Im[G(\br,\br,\omega)],
\eea
where $\psi_n(\br)$ is an eigenfunction of the Hamiltonian, having eigenvalue $E_n$. The Fourier transform of $R(\br)$ is, in terms of the retarded Green's function,
\bea
R(\bq,\omega)=\frac{1}{N\pi}\sum_{\bk}Tr[G(\bk,\bk+\bq,\omega)-G^\ast(\bk+\bq,\bk,\omega)],
\eea
which is identical to the expression of $\rho(\br)$ in the case of a single impurity. But now $G(\bk,\bk+\bq,\omega)$ does not have a simple expression as in Eq.(3) in the text, since the system contains more than one impurity.

By definition, we have the Fourier transform of the correlation function $C(\br)$ as (from here, we suppress the argument of $\omega$ for concision)
\bea
C(\bq)&=&\frac{1}{(2\pi)^3N}\int{d^3r'd^3r}\langle{}R(\br+\br')R(\br')\exp(-i\bq\cdot\br)\rangle\\
\nonumber&=&\langle|R(\bq)|^2\rangle.\eea Therefore the Fourier transform of the autocorrelation function $C(\br)$ is exactly the ensemble average of the module squared Fourier component $R(\bq)$.

For an arbitrary configuration of identical impurities, the retarded Green's function is
\bea
G(\bk,\bk+\bq)=G_0(\bk)\delta_{\bq0}+G_0(\bk)V(\bq)G_0(\bk+\bq)+\frac{1}{N}\sum_{\bk'}G_0(\bk)V(\bk-\bk')G_0(\bk')V(\bk'-\bk-\bq)G_0(\bk+\bq)+...
\eea
where $V(\bq)=V\sum_i\exp(-i\bq\cdot\br_i)$ is the Fourier transform of the impurity potential. From the above equation we have $R(\bq)$ as
\bea
&&R(\bq)=\rho_0\delta_{\bq0}+\frac{1}{iN}\sum_\bk Tr[G_0(\bk)V(\bq)G_0(\bk+\bq)-G^\ast_0(\bk+\bq)V(\bq)G^\ast_0(\bk)]\\
\nonumber&+&\frac{1}{iN^2}\sum_{\bk,\bk'}\sum_\bk Tr[G_0(\bk)V(\bk-\bk')G_0(\bk')V(\bk'-\bk-\bq)G_0(\bk+\bq)-G_0^\ast(\bk+\bq)V(\bk'-\bk-\bq)G^\ast_0(\bk')V(\bk-\bk')G^\ast_0(\bk)]+...
\eea
For $\bq\neq0$, the lowest order in $|\rho(\bq)|^2$ is the second order in $V$:
\bea
|R(\bq)|^2&=&|Tr\{\frac{1}{N}\sum_{\bk}[G_0(\bk)V(\bq)G_0(\bk+\bq)-G^\ast_0(\bk+\bq)V(\bq)G_0^\ast(\bk)]\}|^2|\sum_i\exp(-i\bq\cdot\br_i)|^2\\
\nonumber&=&|\rho(\bq)|^2|\sum_i\exp(-i\bq\cdot\br_i)|^2.
\eea
Finally, we use the independence (randomness) of the distribution of $\br_i$ to prove that
\bea\langle|\sum_i\exp(-i\bq\cdot\br_i)|^2\rangle=N_i+\sum_{i\neq{j}}\langle\exp(i\bq\cdot(\br_j-\br_i))\rangle
\nonumber&=&N_i,
\eea
where $N_i$ is the total number of impurities. Up to this point, we have proved that Fourier transform of the autocorrelation function $C(\br)$ of the overall LDOS $R(\br)$ is proportional to $|\rho(\bq)|^2$ in the limit of weak impurities. In other words, the singularities in $\rho(\bq)$ can be revealed in the autocorrelation of LDOS in the case of multiple identical impurities.

\end{appendix}

\end{document}